\documentclass{ws-ijmpd}
\usepackage{indentfirst}
\usepackage[english]{babel}
\usepackage[mathscr]{eucal}
\usepackage{epsfig}
\usepackage{subfigure}
\usepackage{caption}

\def\lsim{\mathrel{\raise2pt\hbox to 8pt{\raise -5pt\hbox{$\sim$}\hss{$<$}}}}

\begin{document}
\markboth{R. Lisboa, M. Malheiro, A. S. de Castro, P. Alberto and
M. Fiolhais} {Perturbative breaking of the pseudospin symmetry in
the relativistic harmonic oscillator}
%
\catchline{}{}{}{}{}
%
\title{Perturbative breaking of the pseudospin symmetry in the
relativistic harmonic oscillator}
\author{R. LISBOA and M.MALHEIRO}
\address{Instituto de F{\'\i}sica, Universidade Federal
Fluminense, 24210-340 Niter{\'o}i, Brazil\\
ronai@if.uff.br}
\author{A. S. de CASTRO}
\address{Departamento de F{\'\i}sica e Qu{\'\i}mica,
Universidade Estadual Paulista, 12516-410 Guaratinguet\'a, S\~ao
Paulo, Brazil}
\author{P. ALBERTO and M. FIOLHAIS}
\address{Departamento de F{\'\i}sica and Centro de F{\'\i}sica
Computacional, Universidade de Coimbra, P-3004-516 Coimbra,
Portugal}

\maketitle
\begin{history}
\end{history}

\vskip-0.5cm
\begin{abstract}
We show that relativistic mean fields theories with scalar, $S$,
and vector, $V$, quadratic radial potentials can generate a
harmonic oscillator with exact pseudospin symmetry {\it and
positive energy bound states} when $S=-V$. The eigenenergies are
quite different from those of the non-relativistic harmonic
oscillator. We also discuss a mechanism for perturbatively
breaking this symmetry by introducing a tensor potential. Our
results shed light into the intrinsic relativistic nature of the
pseudospin symmetry, which might be important in high density
systems such as neutron stars.
\end{abstract}

\vskip1cm
Recently, there has been much interest in understanding the
nuclear pseudospin symmetry in terms of relativistic dynamics
\cite{gino}. However, in the presence of nuclear scalar and vector
mean-field potentials this symmetry is broken non-perturbatively
as it has been discussed in recent articles \cite{pmmdm,ronai}. In
this work we will show that for quadratic radial scalar, $S$, and
vector, $V$, potentials with $S=-V$ is possible to break the
pseudospin symmetry by introducing a tensor potential that
preserves the form of the harmonic oscillator central potential
but generates a pseudospin orbit term.

\section{Harmonic oscillator and exact pseudospin symmetry}

 In relativistic mean field theories with scalar and vector
potentials the Dirac Hamiltonian for a fermion (nucleon) of mass
$m$ is
\begin{equation}
H_{\mathrm{D}}=\mbox{\boldmath $\alpha $}\cdot \mbox{\boldmath
$p$}+\beta m+
\frac{1}{2}(1+\beta)\Sigma+\frac{1}{2}(1-\beta)\Delta\ ,
\label{Eq:Dirac}
\end{equation}
where $\mbox{\boldmath $\alpha$}$ and $\beta$ are the usual Dirac
matrices. We have introduced the ``sum" and the ``difference"
potentials defined by $\Sigma=V+S$ and $\Delta=V-S$. When
$\Sigma=0$ or $\Delta=0$, due to the matrix structure of
$\frac{1}{2}(1\pm \beta)$, the Dirac Hamiltonian is invariant
under a SU(2) symmetry \cite{gino}. This is a general feature
which does not depend on the particular form of $S$ and $V$. In
this paper we will consider only the case $\Sigma= 0$ ($S=-V$) and
a harmonic oscillator potential with angular frequency $\omega_1$
for the $\Delta$ potential, \textit{i.e.},
$\Delta=\frac{1}{2}m\omega_{1}^{2}\;r^2$.

A more generalized form of the relativistic harmonic oscillator
where the case $\Delta=0$ is also discussed has been presented
recently \cite{ronaiOH}. Following the details of that paper it is
easy to prove that, for the case $\Sigma=0$, the lower component
of the Dirac spinor satisfies the differential equation
\begin{equation}
\biggl[\frac{\mathrm{d}^2\ }{\mathrm{d} r^2}-\frac{\tilde{l}(\tilde{l}+1)}{%
r^2}-\frac{m(\mathcal{E}-m)}{2} \omega_{1}^{2}r^{2}-(m^{2}-\mathcal{E}^2)%
\biggr]f_\kappa(r)=0.  \label{Eq:D2ordfOFr2}
\end{equation}

As shown in \cite{gino}, $\tilde{l}$ is not the non-relativistic
angular momentum $l$ but rather the pseudo-orbital angular
momentum of the lower component of the Dirac spinor. They are
related by $\tilde\ell=\ell-\kappa/|\kappa|$, where the quantum
number $\kappa$ determines whether spins are parallel or
antiparallel. Pseudospin symmetry is exact when doublets with
$j=\tilde{l}\pm\tilde{s}$ are degenerate, where $\tilde s=s$ is
the pseudospin quantum number.

\begin{figure}[!ht]
\parbox[!hb]{6.2cm}{
\begin{center}
\includegraphics[width=6.2cm]{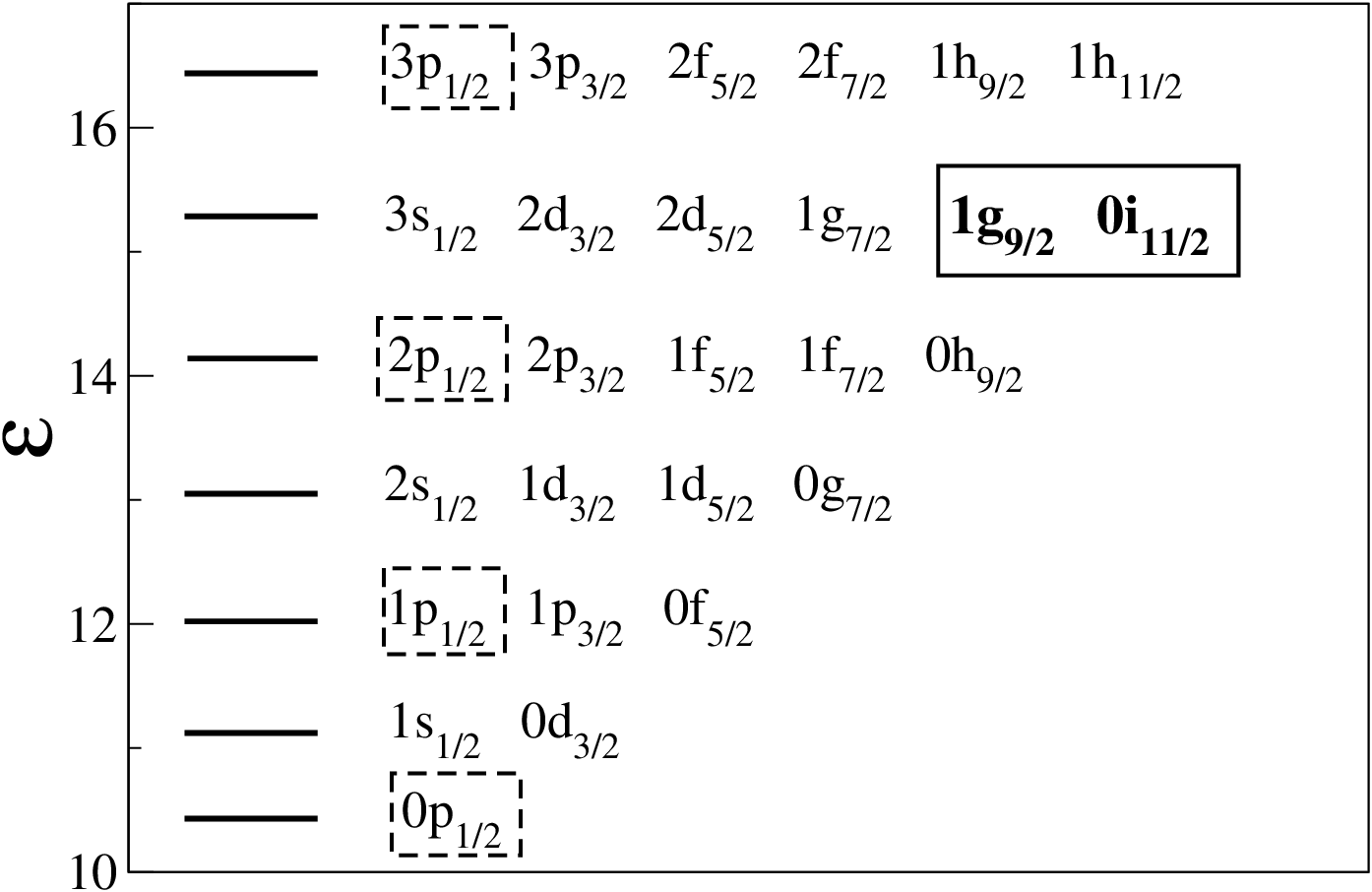}
\end{center}
}\hspace*{.2cm}
\parbox[!hb]{6.2cm}{
\begin{center}
\includegraphics[width=6.2cm]{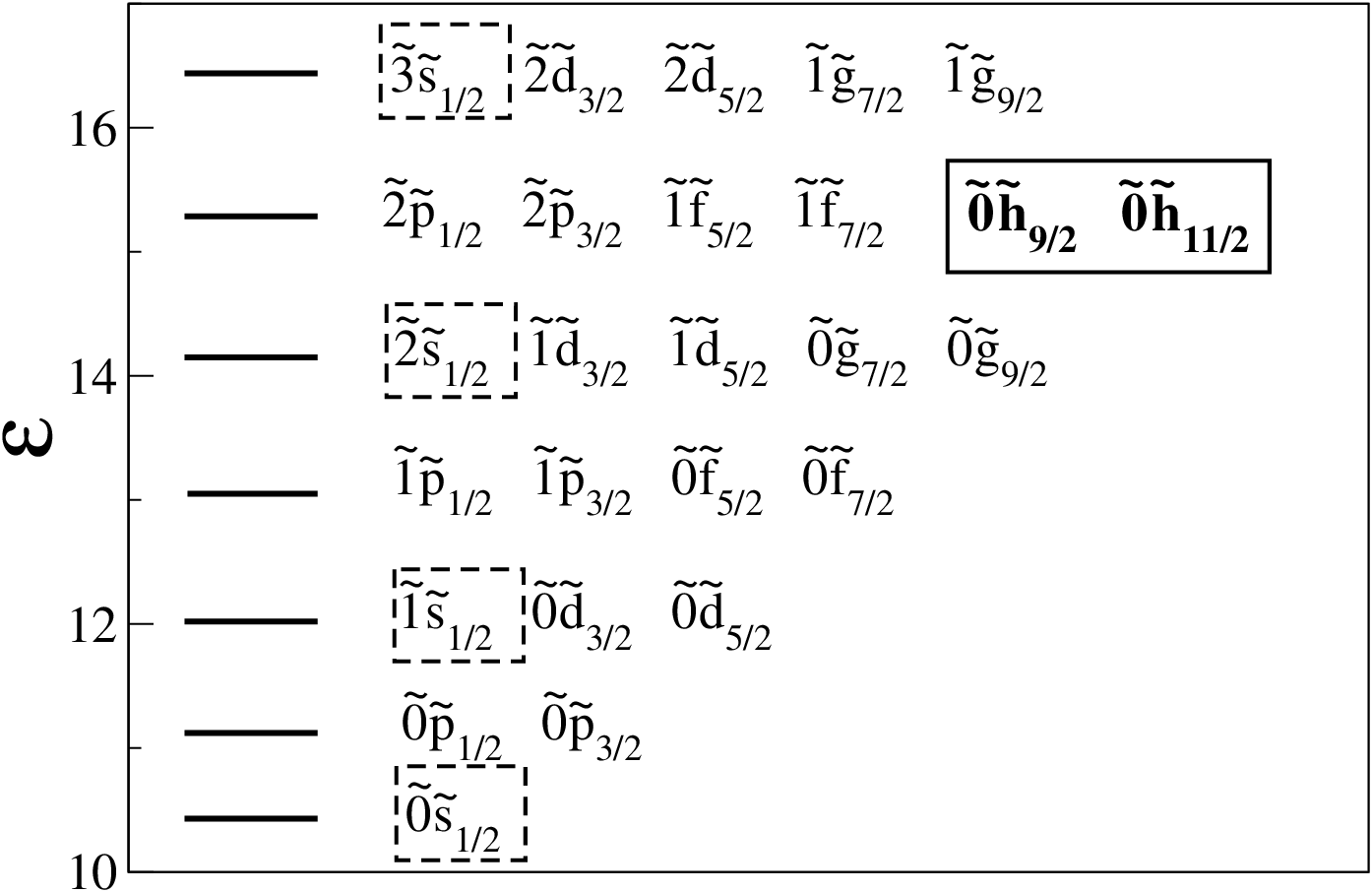}
\end{center}
}\hfill\vspace{-0.2cm} \caption{a) Single particle energies for
the case $\Sigma=0$ with the quantum numbers $(n\,l\,j)$ and b)
$(\tilde{n}\,\tilde{l}\,{\tilde{j}=j})$. The parameters are
$\omega_{1}=2$ and $m=10$.\hfill} \label{Fig:spcPoSignl}
\end{figure}

 The eigenenergies are given by
\begin{equation}
(\mathcal{E}+m)\sqrt{\frac{\mathcal{E}-m}{2m}}=\omega _{1}\left( 2\tilde{n}+%
\tilde{l}+\frac{3}{2}\right) ,\qquad (\tilde{n}=0,1,2,\ldots )\ .
\label{Eq:EnrOHpapertilde}
\end{equation}
where $\tilde{n}$ is the radial quantum number related to the nodes
of the lower component $f_{\kappa}(r)$.
From Eq.~(\ref{Eq:EnrOHpapertilde}) one concludes that the real
solutions must have positive binding energy $E=\mathcal{E}-m$. The
second order equation (\ref{Eq:D2ordfOFr2}), which only depends on
$\tilde{l}$, and also the eigenenergy equation
(\ref{Eq:EnrOHpapertilde}), show that there is no
pseudospin--orbit coupling. Therefore the states with same
$(\tilde{n},\tilde{l})$, but with $j=\tilde{l}+1/2$ and
$j=\tilde{l}-1/2$ are degenerate: they can be seen as pseudospin
doublets as discussed in \cite{gino}.
Thus, when $\Delta$ is a harmonic oscillator potential and
$\Sigma=0$, pseudospin symmetry is exact and there are only
positive-energy bound states, \textit{i.e.}, $\Delta$ acts as a
binding potential. This is an interesting result in view of the
fact that the pseudospin symmetry obtained in the limit
$\Sigma(r)\rightarrow 0$ cannot be realized for nuclear vector and
scalar mean-fields which go to zero as $r\to\infty$. In fact, in
that case, $\Sigma$ acts as a binding negative central potential
well and therefore no bound states may exist when $\Sigma=0$
\cite{gino,pmmdm}. The spectrum of single particle states for the
case $\Sigma=0$, $\omega_{1}=2$ and $ m=10$ is shown in
Fig.~\ref{Fig:spcPoSignl} (a) using the quantum numbers of the
upper component, that can be related to the non-relativistic
quantum numbers.
In Fig.~\ref{Fig:spcPoSignl} (b) we classify the same energy
levels by the quantum numbers of the lower components
$f_{\kappa}$. The comparison between these two figures exhibits
the pseudospin symmetry and its quantum numbers. For example, the
doublets $[1s_{1/2}-0d_{3/2}]$ and $[1p_{3/2}-0f_{5/2}]$, which
have the same pseudo angular momentum $\tilde{l}$ and the same
$\tilde{n}$, are, in the new notation, $[\tilde
0\tilde{p}_{1/2}-\tilde 0\tilde{p}_{3/2}]$ and $[\tilde
0\tilde{d}_{3/2}-\tilde 0\tilde{d}_{5/2}]$, respectively.
Therefore, the harmonic oscillator with $\Sigma=0$ and
$\Delta=\frac{1}{2}m\omega_{1}^{2}r^{2}$  provides an example of
exact pseudospin symmetry. Fig.~\ref{Fig:spcPoSignl} (a) also
shows the $(2\tilde{n}+\tilde{l})$ degeneracy, which means that
not only the states with same $\tilde{n},\tilde{l}$ are degenerate
(pseudospin partners) but also, for example,
$(\tilde{n}-1,\tilde{l}+2)$ or $(\tilde{n}+1,\tilde{l}-2)$ have
the same energy.

The non-relativistic limit of the eigenvalue equation in
Eq.(\ref{Eq:EnrOHpapertilde}) is reached when $\omega_1\ll m$,
which, in turn, means that $E=\mathcal{E}-m\ll 1$, giving
\begin{equation}
E=\frac{1}{2m}\bigg( 2\tilde{n}+\tilde{l}+\frac{3}{2}\bigg)^2 \omega_{1}^{2}\ .
 \label{Eq:E-non-rel-U_0}
\end{equation}
This equation, valid for $\Sigma=0$, shows that the energy is of
second order in the ratio $\omega_{1}/m$, meaning that the energy
is zero up to first order in $\omega_{1}/m$. We can interpret this
fact by saying that, up to this order, there is no
non-relativistic limit for $\Sigma=0$ and therefore the theory is
intrinsically relativistic and so is the pseudospin symmetry.
\vspace{-0.2cm}

\section{Perturbative breaking of the pseudospin by a tensor potential}

When $\Sigma\neq0$ ($S\not=-V$) a pseudospin orbit term shows up
in the equation for the lower component given by
\begin{eqnarray}
-\frac{\Sigma^\prime}{E-\Sigma (r)}\frac{\kappa }{r},
\label{Eq:psterm}
\end{eqnarray}
where the prime denote the derivative with respect to $r$ and
$\kappa$ is the quantum number referred to before. So when
$\Sigma^\prime\neq 0$ the denominator, $E-\Sigma(r)$, can become
very small near the pole and the contribution of this term can be
quite significant, as shown in \cite{marcos}. This manifests the
non-perturbative breaking of this symmetry when $\Sigma\neq 0$, as
it is the case for the nucleus \cite{pmmdm}. However, we may
overcome this problem if we keep $\Sigma^\prime=\Sigma=0$ but a
tensor potential $i\beta\mbox{\boldmath $\alpha\cdot\hat{r}$}U$ is
introduced in the Dirac equation. In this case a new term appears
in the spin-orbit term, $-2\kappa U/r$, as shown in
\cite{ronaiOH}. Because of the product $\mbox{\boldmath
$\alpha\cdot\hat{r}$}$, the pole structure in the denominator
discussed before is absent.
\begin{figure}[!ht]
\begin{center}
\includegraphics[width=6.2cm]{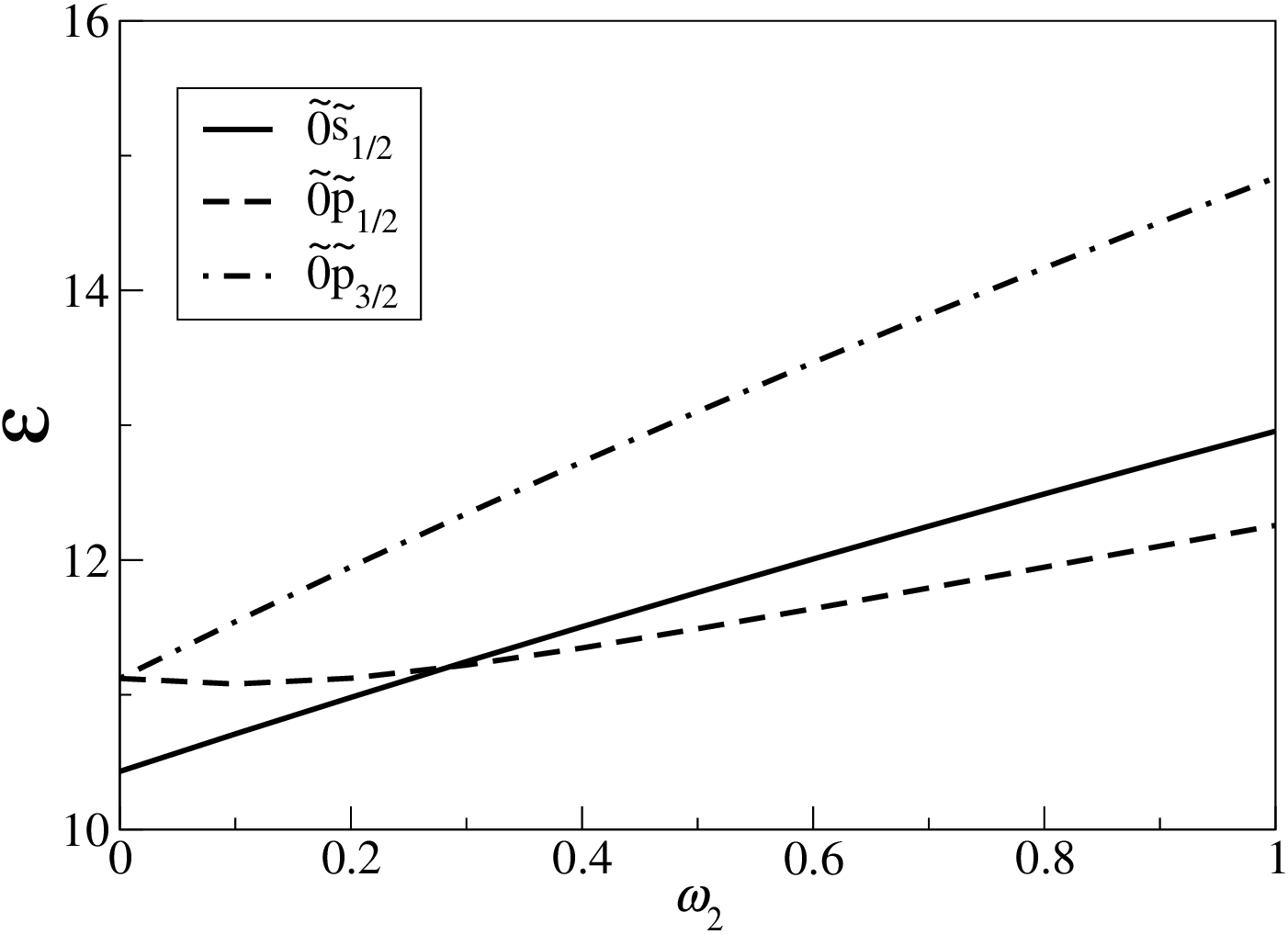}
\end{center}
\caption{Energy levels for $\Sigma=0$
with $\omega_{1}=2$ and $m=10$ as a function of $\omega _{2}$.\hfill}\label{Fig:oh_04}
\end{figure}
\begin{figure}
\begin{center}
\includegraphics[width=6.2cm]{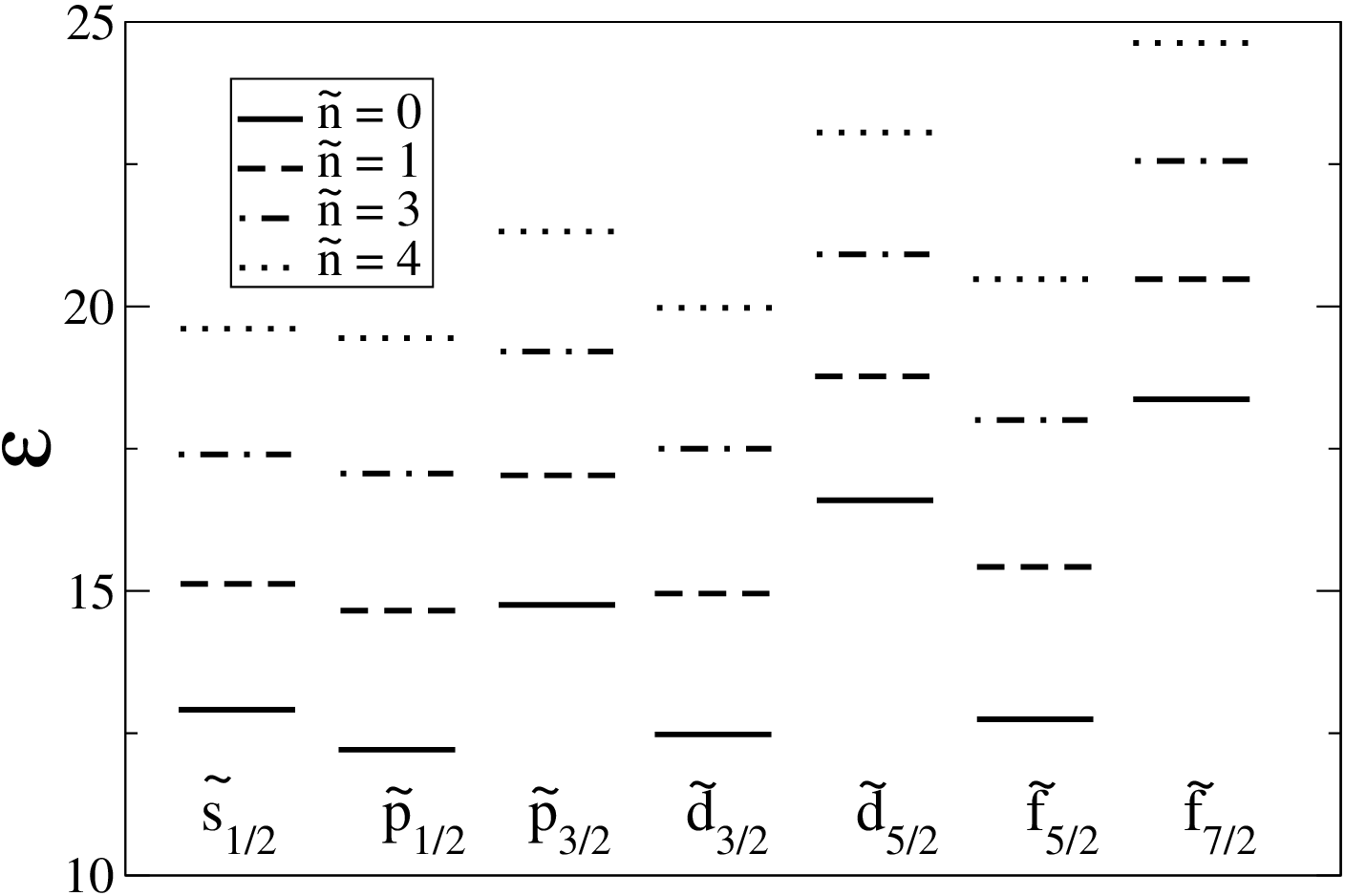}
\end{center}
\caption{Energy spectrum for $\Sigma=0$ with $\omega_{1}=2$, $\omega_{2}=1$
and $m=10$.\hfill} \label{Fig:oh_05}
\end{figure}

If we choose a linear form for $U(r)=m\omega_{2}\,r$, we have an
additional quadratic central potential with a frequency $\omega_2$
in Eq.~(\ref{Eq:D2ordfOFr2}), so that we still have a global
harmonic oscillator central potential. Moreover, we generate a
pseudospin orbit term $-2\kappa\,m\,\omega_2$. Thus, with this new
term, the $\tilde{l}$ degeneracy (pseudospin symmetry) can be
broken perturbatively in the sense that can be made so small as
the $\omega_2$ frequency. We show in Fig.~\ref{Fig:oh_04} how this
symmetry is broken for the $\tilde{p}$ states, for positive
$\omega_2$, and present the eigenenergies in the spectroscopic
notation $\tilde n\,\tilde l_j\,$, where states with
$\tilde{l}=0,1,2,\ldots\,$ are denoted by $\tilde s\,,\tilde
p\,,\tilde d\,,\ldots\,$ respectively. Interestingly, from
Fig.~\ref{Fig:oh_05} we see that with the breaking tensor
potential there is a value of $\omega_2$ for which
$\tilde{0}\tilde{p}_{1/2}$ ($1s_{1/2}$), a state with $l=0$ rather
than $\tilde l=0$, becomes the ground state.

Finally, the relativistic nature of the pseudospin symmetry with
harmonic oscillator potentials shown here suggests that this
symmetry may be more important in ultra-relativistic systems.
Thus, this symmetry may play a role in high density matter, such
as the matter inside compact stars.\vspace{-0.2cm}
\section*{Acknowledgements}\label{agradec}

R. L. and M. M. thanks the nice atmosphere during the IWARA
(Olinda/PE) where this work has been presented.  R. L. and M. M.
acknowledge support from CNPq.  A.S. de C. also thanks support
CNPq and FAPESP.  P.A. and M.F. acknowledges the financial support
from FCT (POCTI), Portugal. \vspace{-0.2cm}

\end{document}